\documentclass[preprint,12pt,3p]{elsarticle}

\usepackage{matlab-prettifier}
\usepackage{lscape}
\usepackage{graphicx}
\usepackage{booktabs}
\usepackage{amsmath}
\usepackage{makecell}
\usepackage{xfrac}
\usepackage{parskip}
\usepackage[backref=page]{hyperref}
\usepackage{adjustbox}
\usepackage{setspace}
\hypersetup{ 
backref=true, 
bookmarks=true, 
pdftoolbar=true, 
pdfmenubar=true, 
pdffitwindow=false, 
pdfstartview={FitH}, 
pdftitle={}, 
pdfauthor={}, 
pdfsubject={}, 
pdfkeywords={}{}, 
pdfnewwindow=true, 
colorlinks=true, 
linkcolor=BlueViolet, 
citecolor=maroon, 
urlcolor=blue,
}
\usepackage{comment}
\usepackage{subfig}
\usepackage{caption}
\usepackage{longtable}
\usepackage{amssymb}
\usepackage{gensymb}
\usepackage{siunitx}

\usepackage{natbib}
\biboptions{comma,round}
\setcitestyle{authoryear}

\begin{document}
\begin{frontmatter}

\title{Climate change, sea level rise, and the agricultural revolution}

\author[label0]{Ridhika Mishra}
\author[label1,label2,label3,label4,label5,label6]{Richard S.J. Tol\corref{cor1}
}
\address[label0]{Indian Institute of Science Education and Research Bhopal, India}
\address[label1]{Department of Economics, University of Sussex, Falmer, United Kingdom}
\address[label2]{Institute for Environmental Studies, Vrije Universiteit, Amsterdam, The Netherlands}
\address[label3]{Department of Spatial Economics, Vrije Universiteit, Amsterdam, The Netherlands}
\address[label4]{Tinbergen Institute, Amsterdam, The Netherlands}
\address[label5]{CESifo, Munich, Germany}
\address[label6]{Payne Institute for Public Policy, Colorado School of Mines, Golden, CO, USA}

\cortext[cor1]{Jubilee Building, BN1 9SL, UK}

\ead{r.tol@sussex.ac.uk}
\ead[url]{http://www.ae-info.org/ae/Member/Tol\_Richard}

\begin{abstract}
We test various hypotheses for the invention and spread of agriculture: higher temperatures, increased seasonality, sea level rise, and technological progress on the extensive margin. We successfully replicate Matranga (QJE 2024). His seasonality hypothesis survives alternative data, estimators, competing hypotheses, and, for adoption but not invention, bootstrap clustering. Population density played a role in the \emph{invention} of agriculture, while spatial diffusion was important for its \emph{spread}. Sea level rise and climate change had a minor effect on the population density of the areas that first adopted agriculture. Extrapolating a logit model for the invention of agriculture, some now-submerged areas in the Persian Gulf, Mediterranean, and Yellow Sea were suitable for early agriculture. Over time, technological progress was the main factor in extending agriculture's extensive margin.
\end{abstract}

\end{frontmatter}

\section{Introduction}
\label{sc:intro}
The Agricultural Revolution, the transition from hunting-gathering to farming, is a key stage in economic development, yet poorly understood. Writing was invented much later, so there are no eyewitness accounts, only myths. These myths tell of people, sometimes from overseas, who came and brought agriculture. Myths also speak of floods that drowned advanced civilizations. We here consider early Holocene sea level rise\textemdash 38 metres between 12,000 and 7,000 years ago\textemdash and its implications for the invention and diffusion of agriculture.

Various hypotheses could explain the agricultural revolution\citep[cf.][]{Weisdorf2005}.\footnote{We do not consider the wider Neolithic Revolution, which includes pastoralism, sedentism, and pottery. \citet{Graeber2021} stress that different populations adopted (and abandoned) elements of the ``Neolithic package'' at different times; and that the transition was gradual rather than revolutionary.} \citet{DowRead2022} and \citet{Olsson2024} argue that \emph{drought} forced people to move to river valleys; overcrowding then called for an intensification of the acquisition of food. \citet{Vince2019} writes that carbon dioxide fertilization stimulated the growth of the grasses that later became cereal crops. \citet{Galor2022} argues it was warmth. He agrees with \citet{ashraf2015} that the relatively stable post-glacial climate allowed for the intergenerational accumulation of farming know-how. Echoing \citet{BranderTaylor1998}, Galor hypothesizes that the population of human hunters rose rapidly after moving into a new area with naive prey. Overhunting led to a crisis and a search for new food sources. \citet{Chu2025} and \citet{Turchin2025} emphasize population pressure \citep[cf.][]{Boserup1965}.\footnote{\citet{Robson2010} argues that agriculture was invented to accommodate a shift in preferences from the quality to the quantity of children.}

Hunting, gathering, fishing and scavenging cannot sustain a dense population; farming and herding can. The rapid loss of coastal lands, typically rich in wildlife, in the early Holocene would have led to population pressures elsewhere, perhaps accelerating the adoption of farming. Combining data on early agriculture with a reconstruction of the climate in the now-submerged continental shelf, we show that farming could have emerged earlier elsewhere.

We test this model against that by \citet{Matranga2024}. He too follows \citet{Boserup1965}. Necessity is the mother of invention, but in Matranga's case hunter-gatherers, he hypothesized, had difficulty coping with the intensification of the seasonal cycle as the ice age ended.

\citet{Matranga2024} is a key data source (see Section \ref{sc:data}), so we start with replicating (successfully) and extending his work in Section \ref{sc:matranga}. Section \ref{sc:submerged} investigates the suitability of now-submerged coastal plains for early agriculture. Section \ref{sc:expansion} studies the impact of Holocene sea level rise and climate change on the diffusion of agriculture. Section \ref{sc:conclusion} concludes.

\section{Data and methods}
\label{sc:data}

\subsection{Data}
\label{sec:data}
Table~\ref{tab:data} summarizes all data sources; the
subsections that follow describe each in turn and explain how it enters the
analysis.

\subsubsection{Historical population and land use: HYDE}
\label{sec:data-hyde}
The population reconstruction takes as its starting point the History Database of the Global Environment (HYDE), a gridded reconstruction, at 5 arc-minute resolution, of population counts and agricultural land use (grazing and crops) back to 12,000~BP. HYDE also includes national and subnational population totals.

\subsubsection{Paleoclimate simulation: TraCE-21ka}
\label{sec:data-trace}
Monthly temperature and precipitation statistics are taken from TraCE-21ka, two transient coupled ocean\textendash atmosphere simulations of the last 22,000 years.

\emph{Archive~I} is the original TraCE-21ka run of \citet{liu2009transient}. \emph{Archive~II} is the revision of \citet{he2022trace}, which substantially reduces meltwater/freshwater forcing during the early-to-middle Holocene (roughly 11,700\textendash 6,000~BP) relative to Archive~I, producing a materially different simulated climate trajectory over that window. Both archives use the same grid: a T31 spectral truncation with a regular 3.75\degree longitude spacing but a non-uniform Gaussian latitude spacing.

From either archive, two seasonality statistics are constructed for a given grid cell and target year, following the definitions of \citet{Matranga2024} and using a 12-month climatology computed over a $\pm15$-year window centered on the nearest available simulation year:
\begin{align}
\text{TempSeas} &= \max(T_{\text{warmest month}}, 0) - \max(T_{\text{coldest month}}, 0), \\
\text{PrecipSeas} &= \frac{P_{\text{wettest month}} - P_{\text{driest month}}}{\overline{P}},
\end{align}
where temperature is $0^\circ$C-censored\textemdash no photosynthesis occurs below freezing regardless of how much colder it gets\textemdash and $\overline{P}$ is the annual mean of all twelve months' precipitation. Mean annual temperature and mean annual precipitation are constructed analogously.

\subsubsection{Modern climate}
\label{sec:data-era5}
We use TraCE-21ka up to 1990~CE. For modern climate, we use ECMWF's ERA5 reanalysis, averaged over 1991\textendash 2020, capping total annual rainfall at 13.5 metres per year. Paleoclimate is constructed from ERA5 and TraCE's simulated change relative to 1990~CE (additive for temperature, multiplicative for precipitation).

\subsubsection{Paleo-topography, ice cover, and sea level}
\label{sec:data-paleo-land}
Which grid cells are dry land and habitable changes over the analyzed period as ice sheets advance and retreat and as global sea level falls and rises. Global bathymetry and land elevation come from GEBCO. Ice-sheet extent comes from \citet{Peltier2015}, supplied as ice-sheet-fraction slices at 0.5~ka time steps; a cell is excluded from the habitable surface if more than 50\% of it is ice-covered. Global mean sea-level relative to the present is taken from \citet{Lambeck2014}, linearly interpolated to any intermediate target year. A cell is classified as land if its GEBCO elevation exceeds that year's sea-level offset. Ruggedness is computed as the within-cell standard deviation of GEBCO elevation at HYDE's native resolution.

\subsubsection{Soil and vegetation}
\label{sec:data-terrain}
Suitability for agriculture is captured by one of two products. The first is the Global Agro-Ecological Zones (GAEZ) soil-quality classification, taking the worst of seven underlying soil factors at each cell. This product describes present-day soil only and does not vary over the reconstruction period. The second, preferred alternative is the terrestrial vegetation and bioclimate reconstruction by \citet{beyer2020}, which supplies a 29-class categorical biome classification, annual net primary productivity, and leaf area index\textemdash all three of which genuinely vary across the reconstruction period. A direct comparison of the two products' ability to classify HYDE's own observed present-day occupancy (Youden's~$J$ statistic) favors the categorical biome classification over both the static soil product and the two continuous vegetation indices; the biome classification is accordingly used as the default terrain-suitability input.

\subsubsection{The invention and diffusion of agriculture}
\label{sec:data-matranga}
Data on the spread of agriculture are taken from \citet{Matranga2024}. Two samples from the replication package are used.

The primary sample is a global panel of 1,090 grid cells on a regular $3.75^\circ \times 3.75^\circ$ grid, combining country-level agricultural adoption dates from \citet{putterman2006} with independent-invention dates and locations from \citet{purugganan2009}. The panel is organized in 500-year periods running from $-19{,}550$ to $1{,}450$~CE. A cell contributes one row per period for as long as it remains ``at risk'' of independently inventing agriculture, exiting the risk set upon either inventing agriculture itself or acquiring it via diffusion from an already-adopting neighbor. Two binary outcome variables are used: an indicator for the seven (cell, period) pairs corresponding to \citet{purugganan2009}'s independent invention events, and a broader indicator for any cell-period at which
agriculture is newly present, whether by independent invention or diffusion.

\citet{Matranga2024}'s climate variables are taken from TraCE-I but regridded to $3.75^\circ \times 3.75^\circ$.



\subsection{Methods}
The methods used are standard. Following \citet{Matranga2024}, we apply ordinary least squares to the cross-sectional data on the date of invention or adoption of agriculture; and panel logit on the same data, restructured as binary data on whether agriculture was \emph{invented} at a particular time and location. We apply panel logit to the question whether agriculture was invented or \emph{adopted}, and make the model dynamic by including previous adoption in nearby locations. We further apply a duration model to the ``hazard'' of exiting hunting and gathering.

We reverse-engineer the algorithm used by the HYDE data to allocate national population numbers to grid cells. That algorithm is used to create counterfactual scenarios: what if sea level had not risen, ice sheets not melted, climate not changed.

\section{Replication and Extension of \citet{Matranga2024}}
\label{sc:matranga}

\citet{Matranga2024} finds that increased climatic seasonality since the Last Glacial Maximum pushed hunter-gatherers toward food storage and sedentism, pre-adapting them to farming. He uses both cross-sectional (equation~7, Table III) and panel data (equation~8, Table IV). We replicate both, using a different paleo-climate reconstruction, and extend the panel specification to include cultural diffusion and population density. We add an eighth invention (New Guinea). We re-estimate the panel as a duration model.

We adopt Matranga's clustering in cells of $15^\circ \times 15^\circ$. As an alternative, because the number of events is small if we consider the independent \emph{invention} of agriculture, we use the wild cluster score bootstrap \citep{Cameron2008, Kline2012}.

\subsection{Cross-sectional replication}
\label{sc:table3}
Matranga's equation~7 regresses a cell's adoption year on temperature and precipitation seasonality measured at that same cell's own adoption year,
\[
Y_i = \alpha + \beta_1 T_i + \beta_2 P_i + \gamma X_i + \varepsilon_i,
\]
where $T_i$ and $P_i$ are the difference between the hottest and coldest months, and the normalized difference between the wettest and driest months, respectively. $X_i$ is a vector of controls (mean temperature, mean precipitation, absolute latitude).  We limit attention to the global sample.

Note that \citet{Matranga2024} regridded TraCE output to a regular $3.75^\circ \times 3.75^\circ$ grid. We maintain the native Gaussian latitude grid. Matranga uses TraCE-I. We use both TraCE-I and TraCE-II. The latter is more realistic.

Table \ref{tab:cross} shows the results. Our estimates are similar to, indeed statistically indistinguishable from Matranga's, almost regardless of the controls included.

\subsection{Panel replication}
\label{sc:table4}
Matranga's equation~8 casts the data as two alternative panels. In both panels, zero denotes that no evidence of agricultural activities was found in this cell for this period. In one panel, one denotes that agriculture was \emph{invented} in that period; Matranga records seven inventions. In the other panel, one denotes that agriculture first appeared in that cell in that period, whether by invention or adoption from elsewhere. In both panels, a cell disappears from the dataset in the period after the appearance of agriculture.

Table~\ref{tab:invention} reproduces (column 1) and replicates (column 2) Matranga's estimates for the invention of agriculture. Replication is not perfect but no parameter estimate is statistically significantly different from his. However, temperature seasonality becomes insignificant if we replace regional clustering with wild bootstrap clustering. Column 3 replaces the earlier paleoclimate reconstruction with the later one. Differences are insignificant. Column 4 restricts the sample to dates after 10,000 BCE so that in Column 5 we can add lagged population density as an explanatory variable. Cells with a greater population density are more likely to invent agriculture (at the 10\% significance level). However, Matranga's hypothesis\textemdash changes in seasonality induced agriculture\textemdash does not lose support (with regional clustering). Column 6, finally, adds the domestication of banana and taro in New Guinea. Again, this does not lead to a statistically significant change in the estimated parameters.

Table \ref{tab:adoption} further extends \citet{Matranga2024}, including not just independent invention but also adoption of agriculture. Matranga does not report results. Column 1 of Table \ref{tab:adoption}, therefore, mirrors column 2 of Table \ref{tab:invention}. The results strengthen Matranga's contention: Stronger seasonality, not just in temperature but also in precipitation, led to faster adoption of agriculture. This is true for either paleoclimate reconstruction (column 2). However, precipitation seasonality becomes statistically insignificant if we add, in column 3, the presence of agriculture in a neighbouring cell in the previous period. Diffusion has a large and highly significant effect. This is still the case if the sample is restricted to 10,000 BCE and later (column 4). Population density is insignificant (column 5) for adoption, unlike invention.

\citet{Matranga2024} applies a logit to duration data, the time spent without agriculture. Table \ref{tab:cox} instead applies Cox' proportional hazards model. A hazard ratio greater than unity implies an earlier exit from hunting-gathering. Column 1 shows the result for invention. Temperature seasonality is positive and significant. Column 2 shows results for the adoption of agriculture. Matranga's seasonality hypothesis is confirmed, also when diffusion (column 3) and population density (column 4) are added. The former is again highly significant but the latter is insignificant.

In sum, \citet{Matranga2024} replicates. His main hypothesis\textemdash increased temperature seasonality stimulated the invention and spread of agriculture\textemdash holds up with alternative data and estimators and when tested against competing explanations.

That said, mean temperature and precipitation are significant factors in some specifications, as is population density. We examine these competing hypotheses in more detail below.

\section{Suitability of Now-Submerged Land for Agriculture}
\label{sc:submerged}

\citet{Matranga2024} is silent about any location that is underwater today. Yet roughly 8 million km$^2$ of land\textemdash the size of Australia\textemdash was dry and ice-free in 10,000 BCE and was drowned by 5,000 BCE. This section uses the paleoclimate and population reconstruction developed in Section~\ref{sc:data} to ask how much of it falls inside the climatic niche that supported early agriculture elsewhere. The next section applies the invention models fitted in Section~\ref{sc:matranga} to estimate the likelihood that agriculture was independently invented on land before it went under.

\begin{landscape}
\begin{figure}
    \centering
    \includegraphics[width=1.0\linewidth]{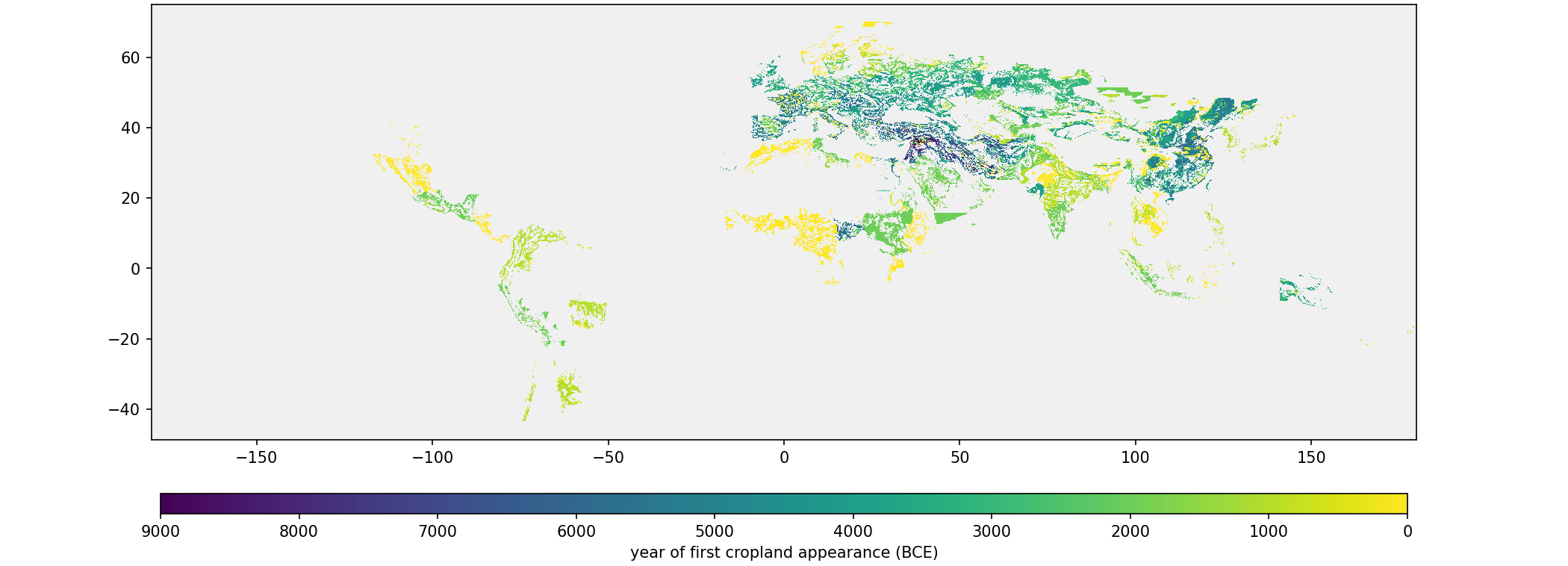}
    \caption{First adoption of agriculture}
    \label{fig:agland}
\end{figure}
\end{landscape}

\begin{figure}
    \centering
    \includegraphics[width=1.0\linewidth]{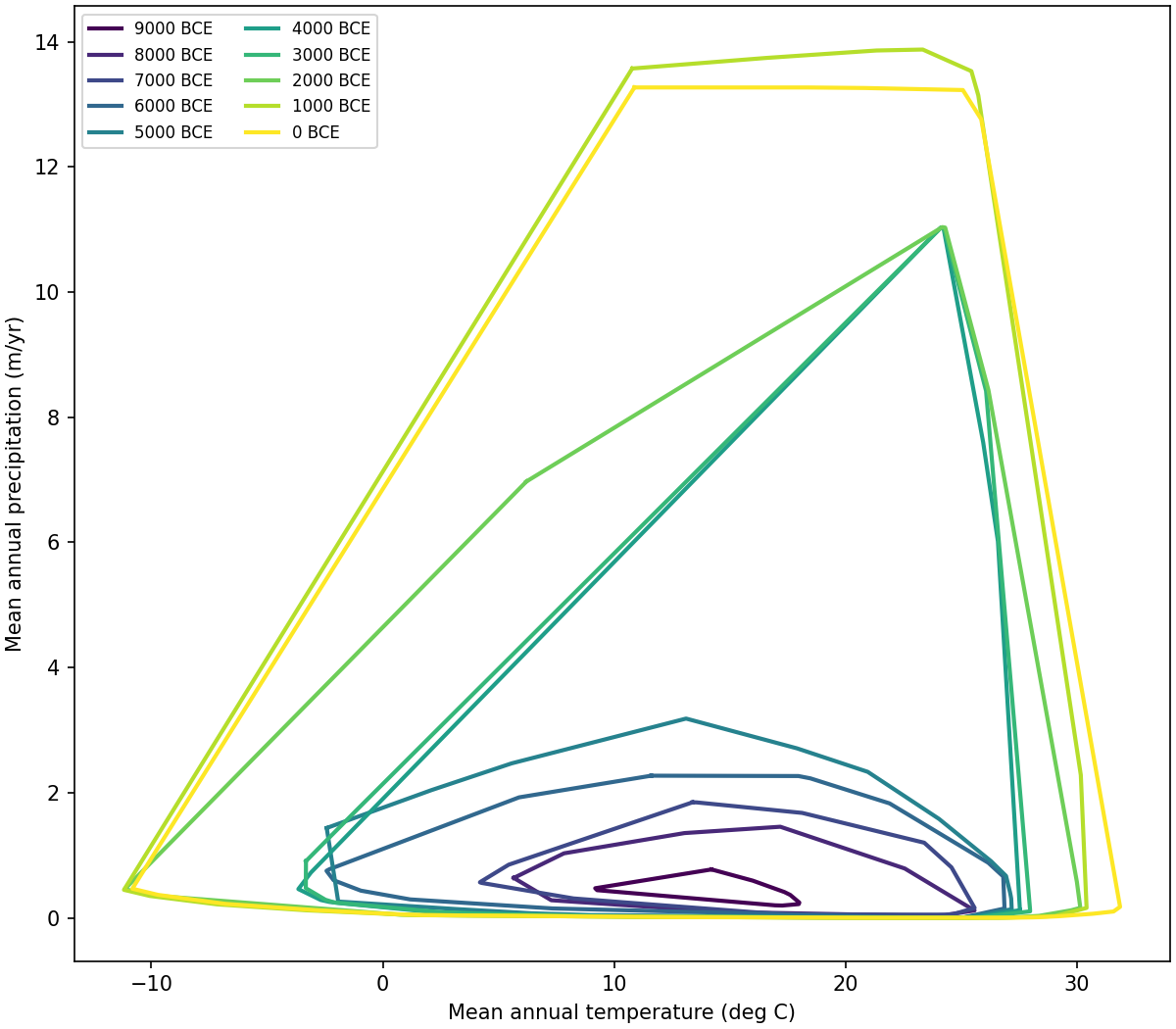}
    \caption{The expanding climate hull of agriculture}
    \label{fig:hull}
\end{figure}

Figure \ref{fig:hull} depicts the climate hull of agriculture. Following \citet{Tol2024niche}, the climate hull is the convex hull in temperature-precipitation space that contains all cells with cropland, according to the HYDE data, in a particular year. The convex hull is a multidimensional generalization of the minimum and maximum. In one dimension, all observations lie between the minimum and maximum. In two (or more) dimensions, all observations lie within the convex hull. Figure \ref{fig:agland} shows how agriculture expanded in geographical space between 9000 and 0 BCE. Figure \ref{fig:hull} shows the same expansion in climate space, after censoring the HYDE data for agriculture at implausibly high elevations (see Table \ref{tab:data}). In 9000 BCE, agriculture was limited to average annual temperatures between 9 and 18\celsius{} and total annual rainfall of 0.2 to 0.8 m/yr. Early agriculture was flood irrigated. By the time of Christ, the temperature range had quintupled while maximum precipitation had increased 17-fold.

Table \ref{tab:niche} shows the area within the climate hull in 5,000 BCE and the counterfactual area, within the climate hull, if sea level had stayed as it was in 10,000 BCE. For this, we use the combined bathymetry and elevation model introduced in Section \ref{sc:data}, only considering low-lying areas connected to open water. We use a separate scenario for the Caspian Sea, and two for the Black Sea, a gradual one (the default) and a catastrophic one (see Table \ref{tab:data}); . We also changed ice cover, but that is irrelevant as ice-covered land lies far outside the climate niche.

For the world as a whole, some 5\% of potentially agricultural land was lost to sea level rise. Regional numbers are higher: around the Mediterranean Sea, 7\% more land would have been arable with early agricultural techniques, 5\% around the Black Sea, and 20\% around the Caspian Sea. The Mediterranean shows the largest \emph{absolute} loss of potential agricultural land. Overall, however, while it is certainly possible that agriculture was first invented and practiced in areas that are now-submerged, sea level rise did not dramatically reduce the acreage of land that was arable with the earliest agricultural methods.

The population maps in the HYDE data combine maps of the population distribution in the recent past with counts and reconstructions of population numbers in jurisdictions in the deeper past, and apply a spatial smoothing algorithm. Unfortunately, the smoothing algorithm is not published. We therefore reverse-engineered it, using temperature, precipitation, ice cover, vegetation, and ruggedness as input variables. Note that, like HYDE, we did not impose any theoretical structure, such as a Malthusian model. Unlike HYDE, we did not maintain borders into the deep past. For example, US states are visible in the HYDE population maps for 10,000 BCE, but not in ours. We then used the same algorithm to construct counterfactual population density map, had sea level and climate stayed as it was in 10,000 BCE.

\begin{landscape}
\begin{figure}
    \centering
    \includegraphics[width=1.0\linewidth]{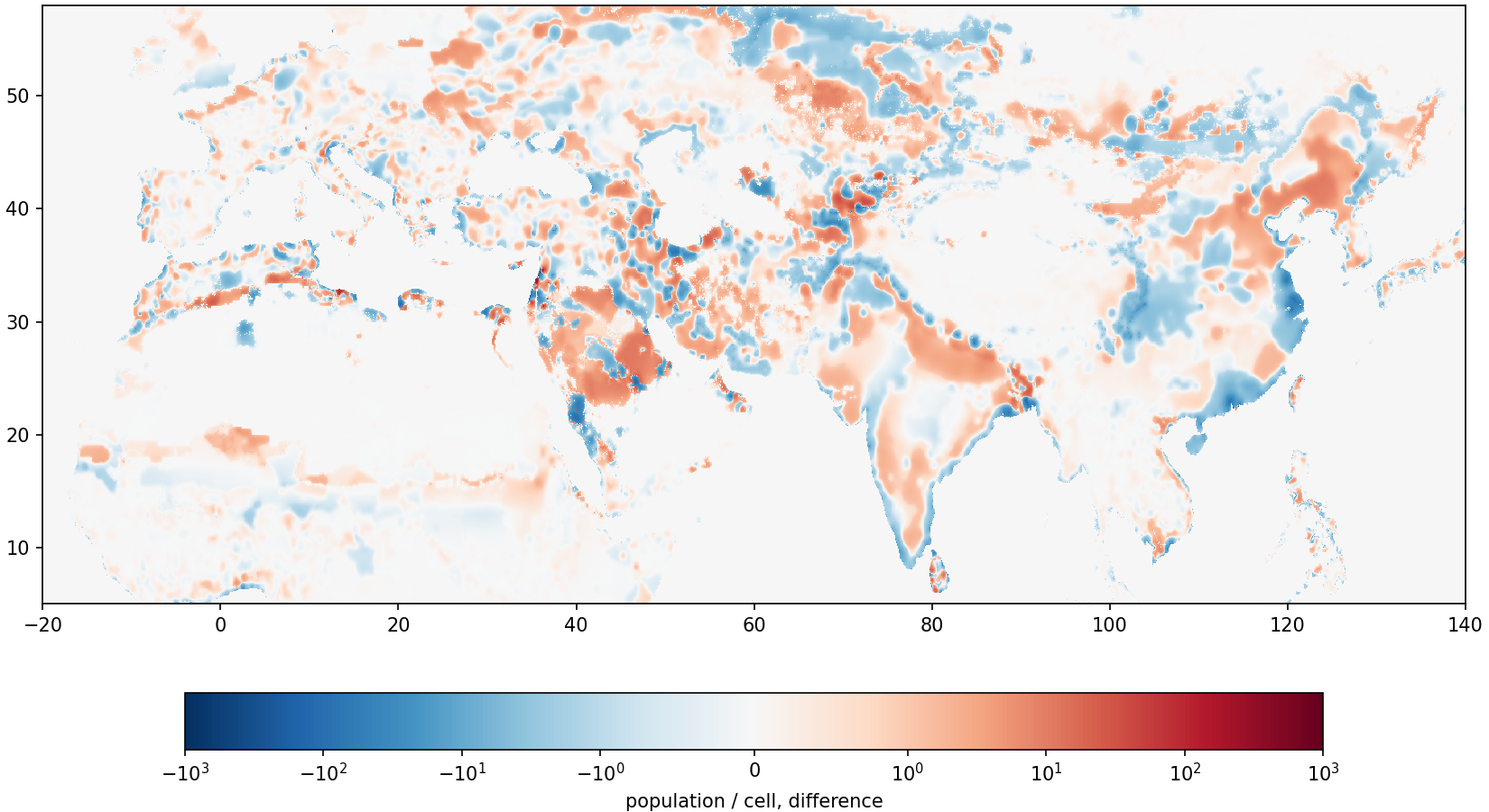}
    \caption{The impact of sea level rise and climate change on population density in 6,000 BCE.}
    \label{fig:popdens}
\end{figure}
    
\end{landscape}

Figure \ref{fig:popdens} shows the results, the difference between actual and counterfactual population density in 6,000 BCE due to sea level rise, climate change, and vegetation change since 10,000 BCE. Evacuation and depopulation of the continental shelf is clearly visible. Adjacent inland areas see an increase in population. We see substantial changes elsewhere too, as previously cold areas warm up and previously wet areas dry out. Changes are particularly large in India and China.

Table \ref{tab:agarea} shows the impact in 6,000 BCE in the areas that had adopted agriculture by 5,000 BCE. Population density decreases by 0.0016 people per square kilometre. This increase is statistically significant, but not substantial: Population density was 0.3207 (0.0022) people per square kilometre, so the increase was not even 0.5\%. Population density was 0.2896 (0.0019) in the areas adjacent to the ones that adopted agriculture first, a difference of about 10\%.

Table \ref{tab:agarea} decomposes the change in the population density, by holding sea level, vegetation, or climate, respectively, as it was in 10,000 BCE. We kept total world population as is in all scenarios. Sea level rise has a \emph{negative} effect. It decreases in the areas that would first adopt agriculture. The mythical crop-bringers preferred to go elsewhere. Climate and vegetation change, however, pushed people towards what later became cropland areas. Changes in vegetation had a somewhat larger impact than climate change but the effects, while statistically significant, are small.

Population density is one of the explanatory variables in the models shown in Table \ref{tab:invention}. Table \ref{tab:submerged} extrapolates these models to the counterfactual, starting with sea level and climate as they were in 10,000 BCE. The cell probability equals $p_c = 1 - \prod_{t=1}^{T} p_{c,t}$, where $p_{c,t}$ is the probability predicted by the logit of Table \ref{tab:invention} and $t$ runs in time-steps of 500 years from 10,000 to $T$, the year of submergence. That is, $p_c$ is the cumulative probability that agriculture would have been invented in cell $c$ \emph{before} it went under.

The top panel of Table \ref{tab:submerged} shows the median cell probability for the world as a whole and 12 regions. The medians are generally low. The bottom panel shows the maximum probabilities. Results differ per model. In our replications of \citet{Matranga2024}, the Yellow Sea emerges as the most likely origin of agriculture, followed by the Tatar Strait and the Red Sea (column B). Using Matranga's coefficients, the Red Sea, Persian Gulf, and Mediterranean appear most likely (columns M). The highest probability, 31\%, is in the Gulf of Suez.

Note the stark difference between the convex hull and logit analyses. The former essentially assigns a zero probability to anything outside the hull. The latter has positive probabilities everywhere. It also uses a richer set of explanatory variables. However, HYDE has only 277 cropland cells in 9,000 BCE. We fitted a 5D convex hull to this (temperature and precipitation mean and seasonality plus population density; results not shown). The result is so constrained that no cell apart from the 277 qualifies.

\section{Agricultural Expansion}
\label{sc:expansion}
Section~\ref{sc:submerged} shows the cumulative change in agricultural potential between 10,000 and 5,000 BCE. Let's decompose that by both mechanism and period. Figure \ref{fig:hull} shows the climate hulls for the period 9,000-0 BCE. Table~\ref{tab:expansion} reports the corresponding numbers. The area within the climate hull grew from 5 million square kilometres to 125 million km\textsuperscript{2}. The biggest push was in the ninth millennium BCE; expansion slowed after 4,000 BCE. Expansion was larger towards colder climates than towards hotter climates. As agriculture started in fairly dry areas, movement towards drier areas was limited. Agriculture pushed into wetter areas, particularly in the second and fifth millennia.

Four factors jointly determine the arable area at each date: sea-level rise removing land, ice-sheet retreat adding to it, climate change shifting which land falls inside a fixed hull, and the hull itself expanding as farming methods improve. These factors interact: sea level and ice cover only matter for land that would otherwise be climate-suitable. A sequential decomposition is order-dependent, so we instead compute the Shapley value: each factor's marginal contribution averaged over all $4!=24$ possible orderings. Table~\ref{tab:shapley} reports the resulting per-interval decomposition.

Chained over all nine intervals, the total 9,000-BCE-to-0 change of about 120 million km\textsuperscript{2} is attributed to sea level $-3.1\%$, ice melt $+0.0\%$, and climate $+0.8\%$, with the remainder ($+102.4\%$) due to technological improvements on the extensive margin. The negative impact of sea level rise first increases as more potential cropland is exposed and then falls as sea level decelerates. Similarly, the impact of ice cover is small at first and grows as agriculture expands northwards. The impact turns negative after 4,000 BCE as glaciers start growing again. Climate change had a large positive effect in the eighth millennium, and a large negative effect in the fourth. Technological progress on the extensive margin was particularly fast in the first 5,000 years, but then slowed down as almost every possibly suitable climate was already used for agriculture.

In sum, the net effect of sea level rise on ancient agriculture is small. Climate change had a larger effect, but fluctuated between positive and negative. Both effects are dominated by technological progress that expanded the extensive margin.

\section{Discussion and conclusion}
\label{sc:conclusion}
Higher temperatures, changing precipitation, increased seasonality, sea level rise, and technological progress on the extensive margin may all have influenced the invention and spread of agriculture. The seasonality hypothesis by \citet{Matranga2024} survives alternative data, estimators, competing hypotheses, and bootstrap clustering (adoption, not invention). Population density played a role in the \emph{invention} of agriculture, while spatial diffusion was important for its \emph{spread}. Sea level rise and climate change had a minor effect on the population density of the areas that first adopted agriculture. Some now-submerged areas in the Persian Gulf, Mediterranean, and Yellow Sea were suitable for early agriculture based on the estimated logit model for the invention of agriculture. Over time, however, technological progress was the main factor in expanding the extensive margin of agriculture.

Several caveats need to be borne in mind. We take absense of evidence in the archaeological record as evidence of absense of agriculture. The archaeological record is always changing, and skewed towards promising and accessible sites. Some national borders are visible in Figure \ref{fig:agland}. Underwater archaeology is difficult, which makes it hard to prove, let alone disprove our hypothesis of farming on the continental shelf. Our inundation model is based on bathymetry alone. The HYDE data do not extend before 10,000 BCE. Its population density is based on a heuristic, rather than on a model of human behaviour. We do not have reconstructions of past soils, and do not use estimates of the abundance of prey animals. We use a single reconstruction of past vegetation, and two reconstructions of past climate from a single model; we therefore have little idea of the uncertainties about our counterfactual pasts. We do not distinguish between migration and knowledge diffusion in the spread of agriculture. All of these things should be improved upon.

For now, however, it appears that the Matranga hypothesis\textemdash increased seasonality aided in the early development and adoption of farming\textemdash has the highest empirical credentials, if augmented with population pressure and the spread of agricultural technologies. The rapidity with which early farmers mastered hitherto hostile climates is impressive, and perhaps reassuring for the 21st century.

\section*{Acknowledgements}
Pawel Dziewulski, Matt Embrey, Rocco d'Este, Illia Pasichnichenko, and Panu Pelkonen had excellent comments on a previous version of the research. Andrea Matranga helpfully explained aspects of his work. We used Claude for most of the data wrangling and analysis. Claude also wrote an early draft that we had to discard because it was so badly written.

\bibliography{master}
\newpage

\begin{landscape}
\begin{table}[htbp]
\centering
\caption{Data sources}
\label{tab:data}
\begin{tabular}{lllccl}
\toprule
Name & Source & Content & \multicolumn{2}{c}{Resolution} & Coverage \\
& & & space & time & \\
\midrule
HYDE & \citet{Klein2010} & Population, cropland*, grazing land & 5' & 1000 yr & 12,000~BP\textendash 2000~CE \\
TraCE-21ka-I & \citet{liu2009transient} & Temperature, precipitation & T31 & monthly & 22,000~BP\textendash 1990~CE \\
TraCE-21ka-II & \citet{he2022trace} & Temperature, precipitation & T31 & monthly & 22,000~BP\textendash 1990~CE \\
ICE-6G\_C & \citet{Peltier2015} & Ice-sheet fraction & 1\degree & 500 yr & 26,000~BP\textendash present \\
GEBCO & \citet{GEBCO} & Bathymetry, elevation & 15'' & - & Present-day (static) \\
Sea level** & \citet{Lambeck2014} & Sea level & global & 2000 yr & 12,000~BP\textendash present \\
GAEZ & \citet{GAEZ} & Soil quality, terrain & 5' & - & Present-day (static) \\
Ecosystem & \citet{beyer2020} & Biome, NPP, LAI & 0.5\textdegree & 1000 yr & 120,000~BP\textendash present \\
ERA5 & \citet{ERA5} & Temperature, precipitation*** & 0.25\degree & annual & 1991\textendash 2020 \\
Agriculture & \citet{Matranga2024} & Adoption, invention & 3.75\degree & 500 yr & 19,550~BCE \textendash 1,450~CE \\
\bottomrule
\end{tabular}
\caption*{\footnotesize *Elevation limits on agriculture from \citet{dAlpoimGuedes2020, Chen2015science} and \citet{Ritchey2024} for Asia, from \citet{Aldenderfer1998} and \citet{Bruno2003} for South America.\\**Data for the Black Sea from \citet{Ryan1997} (catastrophic scenario) and \citet{Giosan2009} (gradual scenario). Data for the Caspian Sea from the review by \citet{Koriche2022}.\\***Precipitation is capped at 13.5 m/yr.}
\end{table}

\end{landscape}

\newpage

\begin{table}[htbp]
\centering
\caption{Year of adoption}
\label{tab:cross}
\begin{tabular}{lccc}
\toprule
 & Matranga & TraCE-I & TraCE-II \\
\midrule
\multicolumn{4}{l}{\emph{no controls}} \\
temp\_seas & $-129.7^{***}$ & $-111.29^{***}$ & $-115.24^{***}$ \\
 & $(-3.95)$ & $(-3.70)$ & $(-4.34)$ \\
 &  & $[-3.15]$ & $[-3.63]$ \\
precip\_seas & $+73.2$ & $-97.22$ & $-86.23$ \\
 & $(0.25)$ & $(-0.56)$ & $(-0.59)$ \\
 &  & $[-0.56]$ & $[-0.61]$ \\
\addlinespace
\multicolumn{4}{l}{\emph{controls}: mean temperature, mean precipitation, absolute latitude} \\
temp\_seas & $-183.0^{***}$ & $-155.75^{***}$ & $-179.98^{***}$ \\
 & $(-3.73)$ & $(-3.84)$ & $(-5.04)$ \\
 &  & $[-3.44]$ & $[-4.17]$ \\
precip\_seas & $-254.8$ & $-436.09$ & $-412.58^{*}$ \\
 & $(-0.80)$ & $(-1.74)$ & $(-1.89)$ \\
 &  & $[-1.64]$ & $[-1.92]$ \\
\addlinespace
\multicolumn{4}{l}{+ population density (lagged), neighbour\_adopted} \\
temp\_seas &  & $-154.19^{***}$ & $-176.06^{***}$ \\
 &  & $(-3.86)$ & $(-5.00)$ \\
 &  & $[-3.48]$ & $[-4.26]$ \\
precip\_seas &  & $-458.88^{*}$ & $-444.70^{**}$ \\
 &  & $(-1.83)$ & $(-2.07)$ \\
 &  & $[-1.71]$ & $[-2.06]$ \\
\addlinespace
\midrule
$N$ & 587 & 541 & 575 \\
Clusters & 80 & 81 & 83 \\
\bottomrule
\end{tabular}
\caption*{\footnotesize OLS coefficients; dependent variable: year of adoption. Parentheses: $z$, standard errors clustered by $15^\circ\times15^\circ$ block (geogroup15, as in Matranga). Square brackets: equivalent $z$ from a wild cluster score bootstrap (Rademacher weights, null imposed). Stars from the bootstrap $p$-value (Matranga's column: from his $z$): $*/**/***$: $p<0.10/0.05/0.01$. Every coefficient in this table is statistically indistinguishable from Matranga's.}
\end{table}

\begin{table}[htbp]
\centering
\caption{Invention of agriculture}
\label{tab:invention}
\begin{adjustbox}{max width=\textwidth}
\begin{tabular}{lcccccc}
\toprule
 & (1) & (2) & (3) & (4) & (5) & (6) \\
 & Matranga & TraCE-I & TraCE-II & TraCE-II & Population & NG \\
\midrule
temp\_seas & $1.189^{**}$ & $1.160$ & $1.153$ & $1.158$ & $1.156$ & $1.147$ \\
 & $(2.48)$ & $(2.18)$ & $(2.22)$ & $(2.17)$ & $(2.07)$ & $(2.17)$ \\
 &  & $[1.46]$ & $[1.49]$ & $[1.47]$ & $[1.41]$ & $[1.50]$ \\
precip\_seas & $1.854$ & $1.245$ & $1.495$ & $1.325$ & $1.326$ & $1.238$ \\
 & $(1.03)$ & $(0.42)$ & $(1.12)$ & $(0.83)$ & $(0.75)$ & $(0.54)$ \\
 &  & $[0.25]$ & $[0.60]$ & $[0.47]$ & $[0.40]$ & $[0.30]$ \\
mean\_temp & $0.992$ & $0.949$ & $0.943$ & $0.854$ & $0.862$ & $0.915$ \\
 & $(-0.12)$ & $(-0.59)$ & $(-0.70)$ & $(-1.65)$ & $(-1.42)$ & $(-0.92)$ \\
 &  & $[-0.49]$ & $[-0.53]$ & $[-1.33]$ & $[-1.11]$ & $[-0.78]$ \\
mean\_precip & $1.851^{**}$ & $1.539^{**}$ & $1.592^{*}$ & $1.336$ & $1.328$ & $1.639$ \\
 & $(2.15)$ & $(1.83)$ & $(1.77)$ & $(1.00)$ & $(0.91)$ & $(1.74)$ \\
 &  & $[2.13]$ & $[1.78]$ & $[0.89]$ & $[0.85]$ & $[1.30]$ \\
abs\_lat & $1.025$ & $0.991$ & $0.994$ & $0.924$ & $0.929$ & $0.954$ \\
 & $(0.72)$ & $(-0.15)$ & $(-0.10)$ & $(-1.25)$ & $(-1.05)$ & $(-0.76)$ \\
 &  & $[-0.12]$ & $[-0.07]$ & $[-0.99]$ & $[-0.74]$ & $[-0.61]$ \\
ln(PD).L1 & \textemdash & \textemdash & \textemdash & \textemdash & $7.927^{*}$ & $7.411^{*}$ \\
 &  &  &  &  & $(6.23)$ & $(6.18)$ \\
 &  &  &  &  & $[1.87]$ & $[1.90]$ \\
\midrule
N & 16,989 & 17,104 & 17,545 & 8,308 & 8,308 & 8,320 \\
\#events & 7 & 7 & 7 & 7 & 7 & 8 \\
Clusters &  80 & 81 & 87 & 86 & 86 & 87  \\
Pseudo-$R^2$ & 0.080 & 0.062 & 0.068 & 0.076 & 0.096 & 0.077 \\
\bottomrule
\end{tabular}
\end{adjustbox}
\caption*{\footnotesize Odds ratios. Parentheses: $z$, standard errors clustered by $15^\circ\times15^\circ$ block (as in Matranga). Square brackets: equivalent $z$ from a wild cluster score bootstrap (Rademacher weights, null imposed). Stars from the bootstrap $p$-value: $*/**/***$: $p<0.10/0.05/0.01$. Column (1): Matranga's published estimates (stars from his $z$); number of clusters are taken from the replication package, not the paper. Column (4): sample with population data. NG: New Guinea added as an eighth invention.}
\end{table}

\begin{table}[htbp]
\centering
\caption{Adoption of agriculture}
\label{tab:adoption}
\begin{adjustbox}{max width=\textwidth}
\begin{tabular}{lccccc}
\toprule
 & (1) & (2) & (3) & (4) & (5) \\
 & TraCE-I & TraCE-II & TraCE-II & TraCE-II & TraCE-II \\
\midrule
temp\_seas & $1.027^{***}$ & $1.036^{***}$ & $1.033^{***}$ & $1.026^{***}$ & $1.026^{***}$ \\
 & $(3.35)$ & $(4.36)$ & $(3.25)$ & $(2.67)$ & $(2.66)$ \\
 & $[3.34]$ & $[4.42]$ & $[3.16]$ & $[2.67]$ & $[2.66]$ \\
precip\_seas & $1.185^{***}$ & $1.156^{***}$ & $1.080$ & $1.056$ & $1.058$ \\
 & $(3.44)$ & $(3.24)$ & $(1.36)$ & $(1.06)$ & $(1.07)$ \\
 & $[3.06]$ & $[2.80]$ & $[1.32]$ & $[1.03]$ & $[1.05]$ \\
mean\_temp & $1.098^{***}$ & $1.101^{***}$ & $1.071^{***}$ & $1.019$ & $1.019$ \\
 & $(5.35)$ & $(5.24)$ & $(3.42)$ & $(0.95)$ & $(0.94)$ \\
 & $[4.42]$ & $[4.42]$ & $[3.92]$ & $[0.97]$ & $[0.96]$ \\
mean\_precip & $1.283^{***}$ & $1.305^{***}$ & $1.252^{***}$ & $1.168^{***}$ & $1.169^{***}$ \\
 & $(6.04)$ & $(6.48)$ & $(3.86)$ & $(2.77)$ & $(2.78)$ \\
 & $[4.42]$ & $[4.42]$ & $[4.06]$ & $[2.85]$ & $[2.88]$ \\
abs\_lat & $1.069^{***}$ & $1.065^{***}$ & $1.049^{***}$ & $1.020^{**}$ & $1.020^{**}$ \\
 & $(5.84)$ & $(5.95)$ & $(4.51)$ & $(1.95)$ & $(1.95)$ \\
 & $[4.42]$ & $[4.42]$ & $[4.42]$ & $[1.96]$ & $[1.97]$ \\
neighbor\_adopted & \textemdash & \textemdash & $21.305^{***}$ & $9.955^{***}$ & $9.799^{***}$ \\
 &  &  & $(16.65)$ & $(11.62)$ & $(10.49)$ \\
 &  &  & $[4.42]$ & $[4.42]$ & $[4.42]$ \\
ln(PD).L1 & \textemdash & \textemdash & \textemdash & \textemdash & $1.120$ \\
 &  &  &  &  & $(0.31)$ \\
 &  &  &  &  & $[0.31]$ \\
\midrule
N & 17,104 & 17,545 & 17,545 & 8,308 & 8,308 \\
\#events & 538 & 569 & 569 & 569 & 569 \\
Clusters & 81  & 87  & 87  & 86  & 86 \\
Pseudo-$R^2$ & 0.020 & 0.024 & 0.205 & 0.140 & 0.140 \\
\bottomrule
\end{tabular}
\end{adjustbox}
\caption*{\footnotesize Odds ratios. Parentheses: $z$, standard errors clustered by $15^\circ\times15^\circ$ block (as in Matranga). Square brackets: equivalent $z$ from a wild cluster score bootstrap (Rademacher weights, null imposed). Stars from the bootstrap $p$-value: $*/**/***$: $p<0.10/0.05/0.01$. Columns (4)\textendash(5): sample restricted to 10,000~BCE and later.}
\end{table}

\begin{table}[htbp]
\centering
\caption{Cox proportional hazards}
\label{tab:cox}
\begin{tabular}{lcccc}
\toprule
 & (1) & (2) & (3) & (4) \\
 & Invention & Adoption & Adoption & Adoption \\
\midrule
temp\_seas & $1.192^{*}$ & $1.067^{***}$ & $1.058^{***}$ & $1.058^{***}$ \\
 & $(2.70)$ & $(4.74)$ & $(4.34)$ & $(4.41)$ \\
 & $[1.72]$ & $[4.42]$ & $[4.42]$ & $[4.42]$ \\
precip\_seas & $1.431$ & $1.206^{**}$ & $1.133^{*}$ & $1.131^{*}$ \\
 & $(1.16)$ & $(2.51)$ & $(1.94)$ & $(1.90)$ \\
 & $[0.56]$ & $[2.17]$ & $[1.92]$ & $[1.90]$ \\
mean\_temp & $0.836$ & $0.993$ & $1.002$ & $1.001$ \\
 & $(-1.52)$ & $(-0.28)$ & $(0.07)$ & $(0.02)$ \\
 & $[-1.24]$ & $[-0.27]$ & $[0.07]$ & $[0.02]$ \\
mean\_precip & $1.405$ & $1.412^{***}$ & $1.314^{***}$ & $1.313^{***}$ \\
 & $(1.29)$ & $(5.03)$ & $(4.55)$ & $(4.56)$ \\
 & $[1.21]$ & $[4.42]$ & $[4.26]$ & $[4.26]$ \\
abs\_lat & $0.915$ & $1.011$ & $1.014$ & $1.013$ \\
 & $(-1.15)$ & $(0.68)$ & $(1.04)$ & $(0.99)$ \\
 & $[-0.86]$ & $[0.68]$ & $[1.04]$ & $[0.99]$ \\
neighbor\_adopted & \textemdash & \textemdash & $3.712^{***}$ & $3.818^{***}$ \\
 &  &  & $(8.03)$ & $(7.99)$ \\
 &  &  & $[4.42]$ & $[4.42]$ \\
ln(PD).L1 & \textemdash & \textemdash & \textemdash & $0.849$ \\
 &  &  &  & $(-0.52)$ \\
 &  &  &  & $[-0.38]$ \\
\midrule
N & 17,545 & 8,308 & 8,308 & 8,308 \\
\#events & 7 & 569 & 569 & 569 \\
Clusters & 87 & 86 & 86 & 86 \\
\bottomrule
\end{tabular}
\caption*{\footnotesize Hazard ratios. Parentheses: $z$, standard errors clustered by $15^\circ\times15^\circ$ block (geogroup15). Square brackets: equivalent $z$ from a wild cluster score bootstrap (Rademacher weights, null imposed). Stars from the bootstrap $p$-value: $*/**/***$: $p<0.10/0.05/0.01$. TraCE-II. Columns (2)\textendash(4): sample restricted to 10,000~BCE and later.}
\end{table}

\begin{table}[htbp]
\centering
\caption{The impact of sea level rise and climate change on population density in 6,000 BCE in the areas that had adopted agriculture by 5,000 BCE.}
\label{tab:agarea}
\begin{tabular}{lrr}
\toprule
Scenario & \multicolumn{2}{c}{$\Delta$ population density} \\
\midrule
All change & $-0.0016$ & $(0.0004)$ \\
No sea level rise & $-0.0041$ & $(0.0003)$ \\
No vegetation change & $+0.0023$ & $(0.0003)$ \\
No climate change & $+0.0017$ & $(0.0003)$ \\
\bottomrule
\end{tabular}
\caption*{\footnotesize People per km$^2$; standard error in parentheses. 53,206 cells. Each row compares the reconstruction with all changes since 10,000~BCE to the counterfactual named; total world population is the same in all scenarios.}
\end{table}

\begin{table}[htbp]
\centering
\caption{Area (km\textsuperscript{2}) with climate suitable for agriculture (9,000 BCE technology) in 5,000 BCE with and without sea level rise.}
\label{tab:niche}
\begin{tabular}{lrrr}
\toprule
Region & \multicolumn{2}{c}{sea level} & change \\
& \multicolumn{1}{c}{rise} & \multicolumn{1}{c}{constant} & \\
\midrule
Global & 5,218,710 & 5,484,070 & $+5.1\%$ \\
\addlinespace
Mediterranean & 1,027,187 & 1,098,473 & $+6.9\%$ \\
Black Sea (gradual) & 34,123 & 36,008 & $+5.5\%$ \\
Black Sea (flood) & 34,123 & 49,144 & $+44.0\%$ \\
Caspian Sea & 93,439 & 111,803 & $+19.7\%$ \\
\bottomrule
\end{tabular}
\caption*{\footnotesize Sea level constant: sea and lake levels as in 10,000~BCE. Regions: Mediterranean 30\textendash46\degree N, 6\degree W\textendash38\degree E; Black Sea 40\textendash48\degree N, 27\textendash42\degree E, where the two overlap the cell goes to the basin whose present-day sea is nearer; Caspian 36\textendash48\degree N, 46\textendash57\degree E. Not shown are the Persian Gulf, Red Sea, Sunda Shelf, Sahul, Nouadhibou (all too hot), Beringia, Tatar Strait, Doggerland (all too cold) and the Yellow Sea Plain (parts too wet, other parts too cold). Black Sea before reconnection with the Mediterranean (7,450~BCE): gradual, about $-30$~m \citep{Giosan2009}, or flood, about $-100$~m \citep{Ryan1997}. Other rows: gradual scenario.}
\end{table}

\begin{table}[htbp]
\centering
\caption{Predicted invention probability, global and by submerged region}
\label{tab:submerged}
\begin{tabular}{lrccccc}
\toprule
Region & Area (km$^2$) & (A) & (B) & (C) & (M-I) & (M-II) \\
\midrule
\multicolumn{7}{l}{\textit{Panel A: Median probability}} \\
\midrule
Global & 8,186,952 & 0.15\% & 0.18\% & 0.12\% & 0.02\% & 0.03\% \\
Persian Gulf & 116,856 & 0.48\% & 0.52\% & 0.40\% & 1.23\% & 0.90\% \\
Doggerland & 207,322 & 0.16\% & 0.17\% & 0.12\% & 0.03\% & 0.05\% \\
Sunda Shelf & 976,533 & 0.27\% & 0.25\% & 0.18\% & 0.00\% & 0.00\% \\
Sahul & 523,449 & 0.10\% & 0.09\% & 0.07\% & 0.01\% & 0.00\% \\
Beringia & 748,179 & 0.14\% & 0.20\% & 0.13\% & 0.01\% & 0.02\% \\
Mediterranean & 137,588 & 0.16\% & 0.13\% & 0.10\% & 0.16\% & 0.14\% \\
Black Sea (gradual) & 26,031 & 0.48\% & 0.42\% & 0.42\% & 0.46\% & 0.47\% \\
Black Sea (flood) & 88,044 & 0.38\% & 0.34\% & 0.31\% & 0.34\% & 0.35\% \\
Caspian Sea & 276,204 & 0.95\% & 0.78\% & 0.57\% & 1.04\% & 1.03\% \\
Red Sea & 86,070 & 0.29\% & 0.64\% & 0.44\% & 0.20\% & 0.50\% \\
Yellow Sea Plain & 219,735 & 3.13\% & 4.06\% & 3.22\% & 0.50\% & 0.62\% \\
Tatar Strait & 18,750 & 0.97\% & 0.83\% & 0.53\% & 0.06\% & 0.06\% \\
Nouadhibou & 16,500 & 0.08\% & 0.08\% & 0.06\% & 0.15\% & 0.10\% \\
Other shelf & 4,833,735 & 0.13\% & 0.15\% & 0.10\% & 0.02\% & 0.03\% \\
\midrule
\multicolumn{7}{l}{\textit{Panel B: Maximum probability}} \\
\midrule
Global & 8,186,952 & 8.76\% & 8.80\% & 8.26\% & 8.20\% & 30.93\% \\
Persian Gulf & 116,856 & 3.78\% & 2.72\% & 2.48\% & 8.20\% & 6.32\% \\
Doggerland & 207,322 & 0.40\% & 0.37\% & 0.30\% & 0.21\% & 0.22\% \\
Sunda Shelf & 976,533 & 1.19\% & 1.01\% & 0.69\% & 0.05\% & 0.04\% \\
Sahul & 523,449 & 0.89\% & 0.78\% & 0.55\% & 0.23\% & 0.16\% \\
Beringia & 748,179 & 1.15\% & 1.06\% & 0.68\% & 0.07\% & 0.15\% \\
Mediterranean & 137,588 & 1.69\% & 1.48\% & 2.35\% & 4.30\% & 4.54\% \\
Black Sea (gradual) & 26,031 & 1.20\% & 0.95\% & 0.88\% & 1.27\% & 1.17\% \\
Black Sea (flood) & 88,044 & 1.20\% & 0.95\% & 0.88\% & 1.27\% & 1.17\% \\
Caspian Sea & 276,204 & 2.52\% & 2.20\% & 1.94\% & 4.30\% & 4.93\% \\
Red Sea & 86,070 & 1.42\% & 3.75\% & 2.89\% & 6.79\% & 30.93\% \\
Yellow Sea Plain & 219,735 & 8.76\% & 8.80\% & 8.26\% & 1.10\% & 1.19\% \\
Tatar Strait & 18,750 & 6.35\% & 5.54\% & 3.56\% & 0.39\% & 0.44\% \\
Nouadhibou & 16,500 & 0.54\% & 0.64\% & 0.47\% & 2.91\% & 1.14\% \\
Other shelf & 4,833,735 & 4.09\% & 4.19\% & 3.91\% & 2.47\% & 2.19\% \\
\bottomrule
\end{tabular}
\caption*{\footnotesize Cumulative probability, 10,000\textendash5,500~BCE, that agriculture would have been invented in a cell submerged between 10,000 and 5,000~BCE. (A) TraCE-I and (B) TraCE-II: logit with temp\_seas, precip\_seas, mean\_temp, mean\_precip and abs\_lat, fitted on the population-covered sample (Table~\ref{tab:invention}, column (4), and its TraCE-I analogue); (C) adds ln(PD) (Table~\ref{tab:invention}, column (5)). (M-I), (M-II): Matranga's estimates applied to TraCE-I and TraCE-II. Black Sea before reconnection with the Mediterranean (7,450~BCE): gradual, about $-30$~m \citep{Giosan2009}, or flood, about $-100$~m \citep{Ryan1997}. Other rows: gradual scenario.}
\end{table}

\begin{table}[htbp]
\centering
\caption{Expansion of the climate niche}
\label{tab:expansion}
\begin{tabular}{crrrrr}
\toprule
Period & Area & Coldest & Hottest & Wettest & Driest \\
\midrule
 & (mln km$^2$) & (\celsius) & (\celsius) & (m/yr) & (m/yr) \\
\midrule
9000 BCE & 5.42 & $9.2$ & $18.0$ & 0.77 & 0.197 \\
0 BCE & 124.84 & $-10.8$ & $31.9$ & 13.27 & 0.001 \\
\midrule
$\Delta$/100 yr & (\%) & (\celsius) & (\celsius) & (m/yr) & (m/yr) \\
\midrule
9000$\to$8000 & $+20.94$ & $-0.352$ & $+0.757$ & $+0.068$ & $-0.0157$ \\
8000$\to$7000 & $+2.78$ & $-0.146$ & $+0.000$ & $+0.039$ & $+0.0011$ \\
7000$\to$6000 & $+5.20$ & $-0.663$ & $+0.129$ & $+0.043$ & $-0.0042$ \\
6000$\to$5000 & $+0.95$ & $+0.001$ & $+0.031$ & $+0.091$ & $-0.0005$ \\
5000$\to$4000 & $+2.35$ & $-0.123$ & $+0.035$ & $+0.785$ & $-0.0001$ \\
4000$\to$3000 & $+0.03$ & $+0.032$ & $+0.044$ & $+0.000$ & $-0.0001$ \\
3000$\to$2000 & $+1.28$ & $-0.782$ & $+0.217$ & $+0.001$ & $-0.0000$ \\
2000$\to$1000 & $-0.01$ & $-0.004$ & $+0.027$ & $+0.284$ & $-0.0001$ \\
1000$\to$0 & $-0.01$ & $+0.036$ & $+0.145$ & $-0.061$ & $-0.0001$ \\
\bottomrule
\end{tabular}
\caption*{\footnotesize Climate hull of cells with more than 0.1\% cropland (HYDE), excluding cells whose lowest 10\% of land lies above the period's limit of cultivation (Old World: 2,500~m until 3,250~BCE, 3,000~m in 1,650~BCE, 4,000~m from 1,000~BCE; Americas: 3,400~m until 3,000~BCE, 4,500~m from 2,000~BCE). Area: land within the hull, excluding ice.}
\end{table}

\begin{table}[htbp]
\centering
\caption{Shapley decomposition of the change in area suitable for agriculture (km$^2$, chained)}
\label{tab:shapley}
\begin{adjustbox}{max width=\textwidth}
\begin{tabular}{lrrrrr}
\toprule
Period & Total change & Sea level & Ice melt & Climate & Ext. Margin \\
\midrule
9000$\to$8000 BCE & $+30{,}824{,}600$ & $-148{,}218$ & $+0$ & $+331{,}197$ & $+30{,}641{,}622$ \\
8000$\to$7000 BCE & $+11{,}432{,}696$ & $-404{,}480$ & $+0$ & $+1{,}573{,}282$ & $+10{,}263{,}894$ \\
7000$\to$6000 BCE & $+31{,}481{,}760$ & $-768{,}412$ & $+22{,}441$ & $+352{,}903$ & $+31{,}874{,}827$ \\
6000$\to$5000 BCE & $+7{,}818{,}048$ & $-746{,}570$ & $+2{,}078$ & $-666{,}478$ & $+9{,}229{,}018$ \\
5000$\to$4000 BCE & $+22{,}776{,}048$ & $-718{,}557$ & $+414$ & $+697{,}518$ & $+22{,}796{,}673$ \\
4000$\to$3000 BCE & $+337{,}352$ & $-305{,}366$ & $-30$ & $-924{,}554$ & $+1{,}567{,}302$ \\
3000$\to$2000 BCE & $+14{,}965{,}536$ & $-322{,}828$ & $-3{,}433$ & $-76{,}924$ & $+15{,}368{,}721$ \\
2000$\to$1000 BCE & $-76{,}824$ & $-162{,}448$ & $-1{,}831$ & $-495{,}417$ & $+582{,}872$ \\
1000$\to$0 BCE & $-129{,}792$ & $-144{,}931$ & $-33{,}328$ & $+113{,}484$ & $-65{,}017$ \\
\bottomrule
\end{tabular}
\end{adjustbox}
\end{table}

\end{document}